\documentclass[journal,twoside]{IEEEtran}
\usepackage[utf8]{inputenc}
\usepackage[T1]{fontenc}
\usepackage{graphicx}
\usepackage{amsmath,amssymb}
\usepackage{booktabs}
\usepackage{cite}
\usepackage{hyperref}
\usepackage{siunitx}
\graphicspath{{figures/}}
\usepackage{pgfplots}
\usepackage{algorithm}
\usepackage{algpseudocode}
\usepackage{tikz}
\usepackage{array}
\usepackage{adjustbox}
\usepackage{multirow}
\usepackage[font=footnotesize]{subcaption}
\usepackage[compatibility=false]{caption}
\pgfplotsset{compat=1.18}  
\usepgfplotslibrary{groupplots}
\begin{document}

\title{A Domain-Informed Multi-Objective Framework for EEG Channel Selection in Motor Imagery BCIs}

\author{
Dekka Muni Kumar$^{1}$,~\IEEEmembership{} Dhruba Jyoti Kalita$^{1}$,~\IEEEmembership{} Yogesh Kumar Meena$^{1}$~\IEEEmembership{}%
\thanks{$^{1}$ Dekka Muni Kumar, Dhruba Jyoti Kalita and Yogesh Kumar Meena are with Human-AI Interaction (HAIx) Lab, IIT Gandhinagar, India, 
{\tt\small yk.meena@iitgn.ac.in}}}

\maketitle

\pagestyle{headings}


\textbf{\textit{Abstract—} Motor imagery (MI) classification using electroencephalography (EEG) signals is essential for advancing brain-computer interfaces (BCIs). Traditional EEG channel selection methods often face limitations, such as dependency on single-objective criteria and susceptibility to local optima. To address these challenges, this work proposes a multi-objective optimisation framework that employs non-dominated sorting genetic algorithm, multiple-objective particle swarm optimisation, and a multi-objective evolutionary algorithm based on decomposition. Our approach effectively balances spatial relevance—using a Gaussian kernel—and functional discriminability, which assesses intratrial task-related desynchronisation, thereby improving performance. We evaluated this framework on four EEG datasets: Physionet, OpenBMI, HighGamma, and BCIIV-2A. The proposed approach successfully identifies compact, relevant channel subsets concentrated around sensorimotor cortex regions linked to MI activity, addressing the prevalent challenges of dimensionality and complexity inherent to traditional techniques. Furthermore, the framework achieved classification performance of 87\%, 71\%, 75\%, and 65\% on the Physionet, OpenBMI, HighGamma, and BCIIV-2A datasets, respectively. By outperforming existing single-objective and accuracy-based methods, and those relying on fixed subsets, these findings demonstrate that this new multi-objective optimisation framework can enhance MI-based BCI performance while facilitating compact channel configurations with reduced computational complexity, making them better suited for wearable, portable, and real-time BCI applications. }


\section{Introduction}
\IEEEPARstart{M}{otor imagery (MI)} based brain-computer interfaces (BCIs) enable users to control external devices by decoding the electroencephalogram (EEG) signals generated around sensorimotor regions of the brain from the imagined motor movements~\cite{wolpaw2012brain, yuan2014brain}. These systems have a broad range of applications in assistive technologies for paralysed patients~\cite{tarara2025bci, tang2018towards}, neuroprosthetic control of robotic limbs~\cite{exoskeloton, Baniqued2021,mobihealthnews2025}, wheelchair navigation\cite{DBLP:journals/corr/abs-2005-04209, Tang2018}, computer cursor control~\cite{singer-clark2025cursor, forenzo2023integrating}, and virtual reality-based neurorehabilitation therapy for stroke recovery~\cite{Vourvopoulos2016, 10.3389/fnhum.2019.00329}. 

A critical factor for developing effective MI-based BCI systems is optimal EEG channel selection, which ensures the capture of the most relevant neural information while minimising noise and redundancy. Traditional channel selection methods in this context often adopt single-objective approaches or heuristic techniques, such as statistical significance-based selection, filter-based ranking, or domain-knowledge-emphasising sensorimotor regions (e.g., C3, C4) ~\cite{robinson2018neurophysiological, 7318410, 6999180}. Neurophysiological approaches based on hemispheric activation patterns have been explored for channel selection, including lateralisation index (LI)-based methods for identifying contralateral motor cortex activity during MI tasks~\cite{Luo} and generalised lateralised readiness potential (LRP)-based channel optimisation frameworks using non-dominated sorting genetic algorithm (NSGA-II) to obtain compact channel subsets across multiple subjects~\cite{Lee2023}. Handiru et al.~\cite{Vikram} proposed the iterative multi-objective optimisation for channel selection (IMOCS) framework, which iteratively refines channel subsets using spatial proximity and task discriminability measures. Although effective, IMOCS combines multiple objectives into a single aggregated score, limiting exploration of the true Pareto trade-off surface and increasing sensitivity to initialisation. 

To address these limitations, previous works have explored evolutionary multi-objective optimisation techniques. Kee et al.~\cite{Che_Yau_Kee} and Esfahani et al.~\cite{Moein} employed NSGA-II-based frameworks for jointly optimising classification performance and channel reduction. Similarly, Wei et al.~\cite{Wei} and Moubayed et al.~\cite{Al_Moubayed} investigated multiple-objective particle swarm optimisation (MOPSO) and multi-objective evolutionary algorithm based on Decomposition (MOEA/D) strategies for MI channel selection, demonstrating the advantages of Pareto-based optimisation over traditional sequential search methods. More recent studies incorporated multitasking and domain knowledge into the optimisation process. Liu et al.~\cite{Liu} proposed an evolutionary multitasking-based multi-objective optimisation algorithm (EMMOA), while Liu et al.~\cite{dkmoea} introduced a domain-knowledge-assisted multi-objective evolutionary algorithm (DK-MOEA) integrating spatial electrode distance information into the search process.

Traditional channel selection methods in this context often adopt single-objective approaches or heuristic techniques, such as statistical significance-based selection, filter-based ranking, or domain-knowledge-emphasising sensorimotor sites (e.g., C3, C4)~\cite{alotaiby2015review,robinson2018neurophysiological}. While these methods can reduce the number of channels and sometimes improve interpretability, they typically optimise classification accuracy and channel reduction or merge spatial and functional objectives into a single criterion or score. This may bias the selection process, overlooking important trade-offs between noise reduction, anatomical relevance, and signal discriminability ~\cite{dkmoea,amiri2024motor}. As a result, such approaches may include noisy or neurophysiologically irrelevant channels while excluding MI-relevant ones. To address the limitations of traditional channel selection methods, this work proposes a multi-objective optimisation framework. This framework recognises spatial and functional relevance as distinct yet complementary objectives. The goal is to select an optimal subset of channels that is both neurophysiologically meaningful and discriminative in nature~\cite{park2020domain,lux2024selecting}.

\begin{figure*}[!t]
    \centering
    \includegraphics[width=0.85\linewidth]{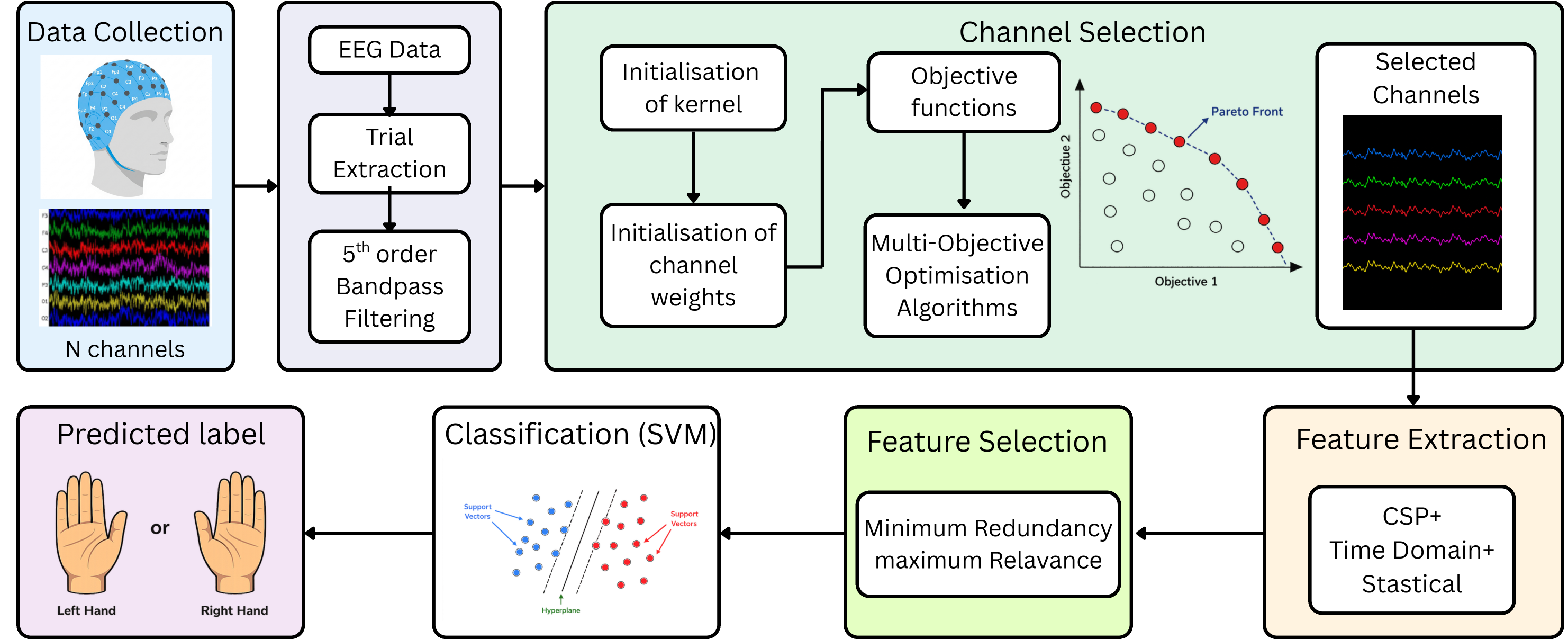}
    \caption{Key components of the proposed domain-informed multi-objective optimisation framework for EEG channel selection.}
    \label{fig:block diagram}
\end{figure*}

To solve the formulated MOO problem, the proposed framework employs algorithms such as NSGA-II~\cite{Deb2002}, MOPSO~\cite{Coello2004}, and MOEA/D~\cite{Zhang2007}, which identify Pareto-optimal solutions (POS) by independently exploring multiple objectives. Unlike hybrid approaches such as IMOCS, which merge criteria into a single score and may obscure important trade-offs, MOO algorithms preserve the independence of objectives, thereby yielding diverse Pareto-optimal channel subsets~\cite{Coello2006}. These population-based global optimisation methods extensively explore the solution space, avoid premature convergence to local optima, and generate multiple alternative solutions~\cite{Deb2002, Zhang2007}. This flexibility enables researchers and clinicians to adapt channel configurations for different analytical scenarios or practical BCI deployment needs. The major contributions of this work are: 

\begin{enumerate}

    \item Proposing a multi-objective optimisation framework for selecting EEG channels in MI classification, combining spatial relevance and functional discriminability with three advanced optimisation algorithms. 
     
    \item Introducing a Gaussian kernel spatial relevance that emphasises electrodes near C3/C4, along with an ITTRD-based functional discriminability to quantify task-related band power changes for EEG channel selection.
    
    \item Implementing optimisation algorithms (i.e., NSGA-II, MOPSO, and MOEA/D)—to preserve independence among the defined objectives, effectively avoiding local optima when finding Pareto-optimal channel subsets.
    
    \item Establishing benchmarks for four public EEG datasets identifies efficient channel subsets in sensorimotor cortex regions linked to motor imagery and enhances classification accuracy, signal quality, interpretability, and computational efficiency. 
    
\end{enumerate}


\section{Proposed Multi-Objective Framework}

The proposed multi-objective framework (see Fig.~\ref{fig:block diagram}) consists of four major stages: preprocessing, multi-objective channel selection, feature extraction and selection, and classification. Initially, raw EEG signals acquired from multiple electrodes are segmented into trials corresponding to MI tasks. The segmented signals are then preprocessed using a 5th-order Butterworth bandpass filter to suppress noise and preserve task-relevant frequency components associated with MI activity. Following preprocessing, channel selection is formulated as a constrained multi-objective optimisation (MOO) problem. In this stage, the kernel is initialised around the sensorimotor reference regions C3 and C4, enabling the assignment of spatial weights to the channels. Subsequently, objective functions corresponding to spatial relevance and functional discriminability are evaluated for the selected channel subsets. To identify Pareto-optimal channel subsets that are both neurophysiologically relevant and effective for MI classification, we employ multi-objective evolutionary algorithms, including NSGA-II, MOPSO, and MOEA/D. 

\subsection{Formulation of Multi-Objective Optimisation Problem}
In this work, the EEG channel selection problem is formulated as a constrained multi-objective optimisation problem presented in Eq.~\ref{eq:moo_problem} that seeks to simultaneously maximise two competing objectives named spatial relevance and functional discriminability~\cite{Vikram}.

\begin{equation}
\begin{aligned}
    & \text{maximize} \quad \{f_{\text{sp}}(x), f_{\text{ITTRD}}(x)\} \\
    & \text{subject to} \quad x \in C, \quad |x| \leq L
\end{aligned}
\label{eq:moo_problem}
\end{equation}

where the decision variable $x$ represents a binary vector encoding a subset of selected channels from the global set $C$, and the maximum limit of channels selected is restricted to L to maintain computational efficiency and practical implementability, where each element indicates whether a particular channel is included (1) or excluded (0) from the optimal solution.

The first objective function in this work is spatial relevance, denoted as \(f_{\text{sp}}\) in Eq.~\ref{eq:moo_problem}, which prioritises the channels located closer to the sensorimotor cortex reference regions C3 and C4. A Gaussian kernel is employed to assign higher weights to channels spatially proximal to the motor cortex while suppressing channels located farther away. The spatial relevance of the \(k\)-th channel is computed as:

\begin{equation}
f_{\text{sp}}^{k} =
\exp\left(
-\frac{
\left\|
\mathbf{C}_{\mathrm{ref}} - \mathbf{C}^{k}
\right\|^{2}
}{
2\sigma^{2}
}
\right)
\label{eq:spatial_relevance}
\end{equation}

where ${f_\text{sp}}^k$ quantifies the spatial relevance between the $k$-th channel and the reference channels $\mathbf{C}_{\mathrm{ref}}$ using a Gaussian kernel. The neighbourhood is bounded by the kernel radius $\sigma$ and is kept as 1 in this work.

The second objective function in this work is intratrial task-related desynchronisation (ITTRD), denoted as \(f_{\text{ITTRD}}\) in Eq.~\ref{eq:moo_problem}, which quantifies the percentage change in spectral power between activation and baseline periods. Unlike conventional ERD/ERS approaches that average spectral power changes across trials~\cite{ERD}, this objective computes the percentage change in spectral power for each individual trial, preserving trial-specific neural characteristics associated with MI tasks. The ITTRD objective is defined as:

\begin{equation}
f_{\text{ITTRD}}(x_i^k)
=
\frac{
\hat{P}_{\text{activation}}
-
\hat{P}_{\text{baseline}}
}{
\hat{P}_{\text{baseline}}
}
\times 100
\label{eq:ittrd}
\end{equation}

where \(\hat{P}_{\text{baseline}}\) and \(\hat{P}_{\text{activation}}\) denote the mean spectral power during the baseline and activation periods, respectively, for the \(i\)-th channel and \(k\)-th trial. The spectral power is estimated using Welch's method as:

\begin{equation}
\hat{P}_{XX}(e^{j\omega})
=
\frac{1}{N}
\left|
\sum_{n=0}^{N-1}
w_R[n]\,x[n]\,e^{-j\omega n}
\right|^2
\label{eq:welch}
\end{equation}

where \(w_R[n]\) denotes the hamming window and \(N\) represents the segment length~\cite{welch}. A negative ITTRD value indicates ERD (power decrease), whereas a positive value indicates ERS (power increase). By jointly optimising \(f_{\text{sp}}\) and \(f_{\text{ITTRD}}\), the proposed framework identifies Pareto-optimal channel subsets that balance anatomical relevance and task-specific neural discriminability for improved MI classification performance. As explained earlier, ${f_{sp}}$ and 
${f_{ITTRD}}$ embody the dual criteria for effective channel 
selection: spatial relevance and the discriminative power of the selected 
channels, respectively. This formulation inherently 
acknowledges the conflicting nature of the objectives, as 
increasing spatial coverage may compromise discriminative power and 
vice versa, necessitating the identification of Pareto-optimal 
solutions that represent optimal trade-offs between these competing criteria.

\subsection{Finding the Optimal Channel Subsets}

The proposed framework employs three multiple-objective optimisation algorithms: NSGA-II, MOPSO, and MOEA/D, to optimise the formulated optimisation problem. These algorithms offer significant advantages for EEG channel selection in MI tasks primarily because of their global, population-based search strategies that effectively explore the complex, high-dimensional search space. NSGA-II excels at maintaining a diverse population of solutions through elitism and crowding distance mechanisms, ensuring that multiple Pareto-optimal channel subsets that balance spatial relevance and functional discriminability are identified. MOPSO, inspired by particle swarm optimisation, on the other hand, utilises collective swarm intelligence to converge rapidly toward optimal solutions while maintaining diversity through an archive of non-dominated solutions. This enables efficient exploration of channel subsets that offer distinct trade-offs and yield better classification accuracy.

MOEA/D decomposes the multi-objective problem into a set of scalar subproblems, which are solved simultaneously by exploiting neighbourhood relations, thereby accelerating convergence to the Pareto-optimal front. Compared to traditional single-objective or aggregated-score methods, which risk bias or local optima, these MOO algorithms independently optimise each objective, producing multiple diverse and interpretable solutions. In this framework, a set of global parameters is used to formalise the optimisation process. $N$ denotes the total number of available channels, and $L$ represents the maximum number of channels that can be selected for the task. The selection process is further guided by a mutation operator that flips each bit from 0 to 1 or vice versa. This operator introduces controlled random variations into the candidate solutions during the MOO and helps maintain diversity within the population, and prevents premature convergence, thereby improving the exploration of the search space. The non-dominated sorting (ND) operator ranks solutions by partitioning the population into successive Pareto-fronts based on dominance relations, and selects the top $P$ solutions.

\begin{equation}
    ND(S) = \{ x \in S \mid \nexists y \in S : F(y) \prec F(x) \}    
\end{equation}

where, $S$ is the population set, $F(x)$ represents objective vector of a solution $x$, and $F(y) \prec F(x)$ indicates that solution $y$ dominates solution $x$.

\subsubsection{Channel selection using NSGA-II}

By implementing the NSGA-II algorithm~\cite{Deb2002} (see Algorithm~\ref{alg:nsga2}), we identify an optimal subset of channels from \( N \) available channels by simultaneously optimising two defined objectives. Each candidate solution is represented as a binary vector $x \in \{0,1\}^N$, where $x_i = 1$ denotes the selection of channel $i$, and $x_i = 0$ denotes its exclusion. To ensure a limited number of selected channels, the algorithm enforces the constraint $\sum_{i=1}^N x_i \leq L$, ensuring that at most $L$ channels are chosen at any iteration. The algorithm starts by randomly initialising $P$ candidate solutions (binary vectors) across the search space. From these candidate solutions, the parent selection is performed via \textit{binary tournament selection}, in which two candidates are randomly sampled, and the one with a superior Pareto rank (or higher crowding distance in case of a tie) is chosen. Offsprings are then generated using \textit{single-point crossover} with probability $p_c$, by selecting a random cut point in the binary vectors and swapping the tail segments between the parents.

Subsequently, the algorithm applies a \textit{bit-flip mutation} to each offspring with probability $p_m$, where each bit is flipped from 0 to 1 or vice versa. A repair step follows to ensure compliance with the $L$-channel constraint. The combined population of parents and offspring is then sorted into non-dominated fronts. From this combined set, the algorithm selects $P$ solutions for the next generation, prioritising Pareto rank and crowding distance to maintain diversity. After $G$ generations, the algorithm outputs the final set of Pareto-optimal solutions, each representing a diverse channel subset that offers distinct trade-offs between the two objectives.


\begin{algorithm}[!t]
\caption{NSGA-II for EEG Channel Selection}
\label{alg:nsga2}
\begin{algorithmic}[1]
\Require 
 $f_{sp}, f_{ITTRD}$: Objectives, $P$: Population size, $G$: No of Generations, 
$p_c$: Crossover prob., $p_m$: Mutation prob.
\Ensure Pareto-optimal channel subsets
\State Initialize $\mathcal{P}^{(0)}=\{\,x^j\in\{0,1\}^N \mid \|x^j\|_1\le L,\ j=1{:}P\,\}$

\For{$g = 1$ to $G$}
    \For{ each $x$ in $\mathcal{P}^{(g-1)}$}
        \State $f_1 = -\sum_i f_{\text{sp}_i} x_i$, \quad $f_2 = -\sum f_{\text{ITTRD}_i} x_i$
        
    \EndFor
    \State 
    $
    \{F_1,F_2,\dots,F_k\}
    =
    ND(\mathcal{P}^{(g-1)})
    $
    
    \State Compute crowding distance for all solutions

    \State Initialise offspring population
    $
    \mathcal{Q}^{(g)}=\emptyset
    $

    \State Initialize offspring set $\mathcal{Q}^{(g)} = \emptyset$
    \For{$j=1$ to $P/2$}
        \State Select $p_1, p_2 \sim \operatorname{BinaryTournament}(\mathcal{P}^{(g-1)})$
       \State $(c_1,c_2) = \operatorname{crossover}(p_1,p_2;p_c)$
        \State $c_1' = \operatorname{mutate}(c_1;p_m), 
        \quad 
        c_2' = \operatorname{mutate}(c_2;p_m)
        $
        \State $\mathcal{Q}^{(g)} \gets \mathcal{Q}^{(g)} \cup \{c_1',c_2'\}$
    \EndFor
    \State  $\mathcal{R}^{(g)} = \mathcal{P}^{(g-1)} \cup \mathcal{Q}^{(g)}$
    \State $\{F_1,F_2,\dots,F_k\} = ND\big(\mathcal{R}^{(g)}\big), 
F_1 \prec F_2 \prec \dots \prec F_k
$
    \State$
\mathcal{P}^{(g)} = \operatorname{Select}(\{F_1,\dots,F_k\}, P)
$

\EndFor
\State \Return $\mathcal{P}^{(G)}$

\end{algorithmic}
\end{algorithm}

\subsubsection{Channel selection using MOPSO} 

To address the formulated MOO problem, we utilise the MOPSO algorithm~\cite{Coello2004} as detailed in Algorithm~\ref{alg:mopso}, where each particle represents a potential subset of EEG channels, encoded as a binary vector. The swarm explores the search space using velocity and position updates, where the velocity controls the probability of flipping a channel’s selection. Each particle tries to improve its own best-known subset ($pbest$) while also learning from the best solutions found by the swarm ($gbest$), which are chosen from the non-dominated repository. The Pareto-based repository ensures that multiple trade-off solutions are maintained rather than a single “best” one. A repair step enforces the maximum channel constraint, and a small-probability mutation helps avoid premature convergence. Over iterations, particles move toward regions of the search space where both objectives are maximised, ultimately producing a diverse set of Pareto-optimal channel subsets.


\begin{algorithm}[!t]
\caption{MOPSO for EEG Channel Selection}
\label{alg:mopso}
\begin{algorithmic}[1]
\Require
$P$: Swarm size, $iter$: No of iterations,  
$w$: Inertia weight, $c_1, c_2$: Acceleration coefficients

\Ensure Pareto-optimal channel subsets

\State Initialise: 
\begin{itemize}
    \item $pos^{(0)} = \{ x^j \in \{0,1\}^N \mid \|x^j\|_1 \leq L,\; j=1,\dots,P \}$
    \item $Vel^{(0)} = 0$
    \item $pbest^{(0)} = pos^{(0)}$
\end{itemize}
\State $Rep^{(0)} = ND(pos^{(0)})$
\State Evaluate $F(x) = \big(-f_{sp}(x), -f_{ITTRD}(x)\big)$
\For{$t=1$ to $iter$}
    \For{$j=1$ to $P$}
        \State Leader: $gbest^j \in Rep^{(t-1)}$ 
        \State update velocity $Vel_j^{(t)}$
        \State $
        pos_j^{(t)} = \sigma(Vel_j^{(t)})   \mid
        \quad \sum\nolimits_{i=1}^N pos_{ji}^{(t)} \leq L$
        
        \State $
        pos_j^{(t)} \gets \operatorname{mutate}(pos_j^{(t)};p_m) 
        $
        
        \State {\footnotesize
$
pbest_j^{(t)} =
\begin{cases}
    pos_j^{(t)}, & \text{if } F(pos_j^{(t)}) \prec F(pbest_j^{(t-1)}) \\
    pbest_j^{(t-1)}, & \text{otherwise}
\end{cases}
$
}
        
    \EndFor
    \State $
    Rep^{(t)} = ND\big(Rep^{(t-1)} \cup pos^{(t)}\big)
    $
\EndFor
\State \Return $Rep^{(iter)}$

\end{algorithmic}
\end{algorithm}


\begin{algorithm}[!t]
\caption{MOEA/D for EEG Channel Selection}
\label{alg:moead-eeg-subproblem}
\begin{algorithmic}[1]
\Require
$P$:No of subproblems; $T$: neighborhood size; $G$: No of generations; $f_{sp},f_{ITTRD}$: Objectives
\Ensure Pareto-optimal channel subsets
\State Define objective: 
$
F(x) = \big(-f_{sp}(x),\,-f_{ITTRD}(x)\big)
$
\State Initialise :
    \begin{itemize}
        \item $
\lambda^k = (\lambda^k_1,\lambda^k_2), \quad  
x^k \in \{0,1\}^N \mid \|x^k\|_1 \le L ,k=1,\dots,P
$
        \item $ B(k) = T\text{ nearest neighbors of }\lambda^k$
        \item Set ideal point $z^* \gets [\min f_{sp}, \min f_{ITTRD}]$
    \end{itemize}
\For{$g = 1$ to $G$} 
    \For{$k = 1$ to $P$} 
        \State Parents: $x^p, x^q \in B(k)$
        \State$
    y \gets \operatorname{mutate}(\operatorname{crossover}(x^p,x^q;p_c);p_m)
    $
        \State Ensure $\|y\|_1 \le L$
         \State Evaluate $
     F(y),\quad z^* = \min(z^*,F(y))
    $
        \For{each $j \in B(k)$}
            \If{$g^{tch}(y,\lambda^j,z^*) < g^{tch}(x^j,\lambda^j,z^*)$}
                \State $x^j \gets y$ 
            \EndIf
        \EndFor
    \EndFor
\EndFor
\State \Return $ND(\{x^1,\dots,x^P\})$
\end{algorithmic}
\end{algorithm}

\subsubsection{ Channel selection uisng MOEA/D}
To identify the optimal channel subsets by simultaneously maximising the two defined objective functions, the MOEA/D algorithm~\cite{Zhang2007} described in Algorithm~\ref{alg:moead-eeg-subproblem} is employed. It decomposes the multi-objective optimisation problem into multiple scalar subproblems, each defined by a unique weight vector representing a specific trade-off between the two objectives. The EEG channel selection was encoded as a binary vector of length $N$, with a constraint of at most $L$ active channels. Each subproblem maintains its own candidate solution and primarily interacts with a neighbourhood of subproblems defined by the Euclidean distance between their weight vectors. Offsprings were generated via single-point crossover and bit-flip mutation with probabilities $P_c$ and $P_m$, respectively, followed by a repair step to satisfy the channel limit. The Tchebycheff aggregation function was used to evaluate and update solutions within neighbourhoods, while the ideal point was dynamically updated as better objective values were found. After $G$ generations, the algorithm produced a set of Pareto-optimal channel subsets that provided diverse trade-offs between the two objectives.

\subsubsection{Channel selection using greedy approach without domain knowledge}

We implemented a greedy channel-selection approach as a baseline for comparison, which does not incorporate domain knowledge~\cite{10.3389/fnins.2022.1045851}. For each subject, the classification accuracy of individual EEG channels is first evaluated using a support vector machine (SVM) with feature selection. Starting from the top-ranked channel, additional channels are appended incrementally in a greedy fashion: at each iteration, the channel that yields the largest accuracy gain or preserves the current best accuracy is added to the selected subset. The procedure terminates when either the subset size reaches sixteen channels or no further improvement is observed. Consequently, the selected subset comprises the channels most relevant to classification for each subject, achieving a favourable balance between accuracy and electrode count. While this strategy may enhance classification performance, it does not guarantee the selection of spatially or neurophysiologically relevant channels~\cite{10650619,10781846}.

\subsection{Feature Extraction and Classification}

After selecting the optimal EEG channel subsets, the filter bank common spatial pattern (FBCSP) was employed to extract spatial-spectral features. Signals from the selected channels were decomposed into nine non-overlapping frequency bands (4-40 Hz), and two CSP spatial filters per class were computed for the binary MI task, resulting in \(9 \times 2 \times 2 = 36\) CSP features per trial~\cite{Vikram}. In addition to CSP features, 19 statistical and time-domain features, including mean, standard deviation, skewness, kurtosis, entropy, Hjorth parameters, zero-crossing rate, Willison amplitude, and slope sign changes, were extracted from each of the \(N_c\) selected channels. For the maximum case of \(N_c = 16\), this resultes in \(19 \times 16 = 304\) statistical features. Consequently, the combined feature space consisted of \(340\) features.

To reduce dimensionality and eliminate redundant information, minimum redundancy maximum relevance (mRMR) was applied to rank features based on mutual information and inter-feature redundancy, retaining only the top 10 features. The selected feature subset was then classified using a support vector machine (SVM) with hyperparameters optimised through grid search.


\section{Performance Evaluation}

The proposed framework is validated on four publicly available EEG datasets for MI classification. The accuracy of each of the three algorithms is determined using features extracted from optimal channel subsets, which are then used to train an SVM classifier. To enhance comparative analysis, a baseline model based on a greedy approach~\cite{10.3389/fnins.2022.1045851} is also implemented without constraints on channel spatial relevance.

\subsection{Datasets and Preprocessing}
This work includes four publicly available EEG datasets, primarily focused on motor imagery (MI)-based brain-computer interfaces (BCIs): 1) Physionet~\cite{physionet}, 2) OpenBMI~\cite{openbmi}, 3) HighGamma~\cite{HGD}, and 4) BCI Competition IV 2A~\cite{bciiv2a}. The PhysioNet dataset consists of EEG recordings from 109 subjects collected using 64 channels at a sampling rate of 160 Hz. It includes left- and right-hand MI trials, with epochs extracted from -0.5 to 4.0 seconds relative to cue onset. The OpenBMI dataset features EEG data from 54 subjects, recorded with 62 channels at 1000 Hz and downsampled to 250 Hz. Session 1 includes 100 trials per class, with epochs spanning from 0 to 8 seconds. The HighGamma dataset contains EEG recordings from 14 subjects, with 128 channels at a sampling rate of 500 Hz, downsampled to 250 Hz. This dataset provides predefined training and test splits for MI tasks. The BCI Competition IV 2A (BCIIV-2A) dataset consists of EEG data from 9 subjects, recorded with 22 channels at a sampling rate of 250 Hz. It includes left and right-hand trials, with epochs ranging from -0.5 to 4 seconds. For all datasets, class labels were binarised (0 = left hand, 1 = right hand). When official train/test splits were available (as in the case of the HighGamma dataset), those were utilised; otherwise, a stratified 80/20 split was implemented to ensure balanced class representation.

Given that EEG datasets vary in recording protocols, channel configurations, and sampling frequencies, preprocessing was carried out to create consistent EEG representations for MI analysis. All EEG signals were filtered using a 5th-order Butterworth bandpass filter within the 4-40 Hz range to preserve the \(\mu\) and \(\beta\) rhythms associated with motor planning and execution while attenuating low-frequency drifts and high-frequency artefacts. Additionally, baseline correction was applied to the Physionet dataset by subtracting the mean signal amplitude computed from the eyes-open resting-state trials. This preprocessing step reduced inter-trial variability and improved the visibility of task-related neural activity.

\begin{table*}
\centering
\caption{Performance comparison of the proposed framework for EEG datasets. ``All'' and ``Sel'' show classification accuracy (\%) for all and selected features, respectively, while "PR" indicates the average number of selected channels (max 16).} 
\label{tab:with_domain_knowledge}
\begin{tabular}{@{}lcccccccccccccc@{}}

\toprule
\multirow{2}{*}{Dataset} & \multirow{2}{*}{\# Ch} & \multirow{2}{*}{\# Sub} & \multicolumn{3}{c}{NSGA-II} & \multicolumn{3}{c}{MOPSO} & \multicolumn{3}{c}{MOEA/D} & \multicolumn{3}{c}{Greedy} \\ \cmidrule(l){4-15} 
 &  &  & All & Sel & PR & All & Sel & PR & All & Sel & PR & All & Sel & PR \\ \midrule
Physionet & 64 & 109 & 73 & 83 & 15.04 & 76 & 87 & 16.00 & 71 & 80 & 15.04 & 91 & 93 & 11.77 \\
BCIIV-2A & 22 & 9 & 52 & 61 & 7.98 & 53 & 63 & 6.31 & 54 & 63 & 10.53 & 52 & 65 & 4.50 \\
HighGamma & 128 & 14 & 70 & 75 & 10.56 & 70 & 74 & 16.00 & 64 & 69 & 13.56 & 67 & 71 & 7.70 \\
OPENBMI & 62 & 54 & 61 & 67 & 10.37 & 61 & 71 & 15.80 & 57 & 54 & 15.70 & 62 & 69 & 4.90 \\ \bottomrule
\end{tabular}
\end{table*}

\begin{figure*}[t]
    \centering
    \begin{subfigure}{0.32\textwidth}
        \centering
        \includegraphics[width=\linewidth]{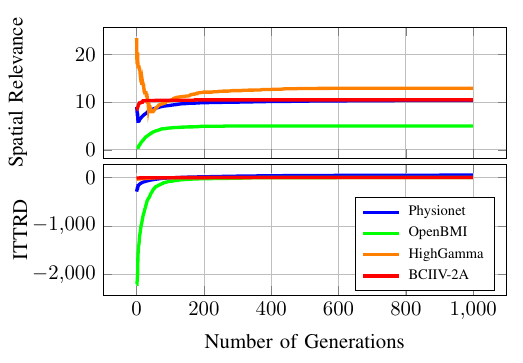}
    \end{subfigure}
    \begin{subfigure}{0.34\textwidth}
        \centering
        \includegraphics[width=\linewidth]{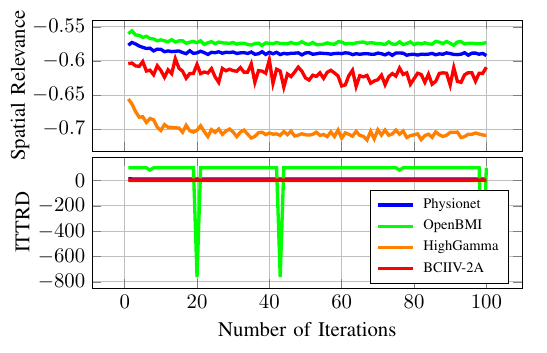}

    \end{subfigure}
    \begin{subfigure}{0.32\textwidth}
        \centering
        \includegraphics[width=\linewidth]{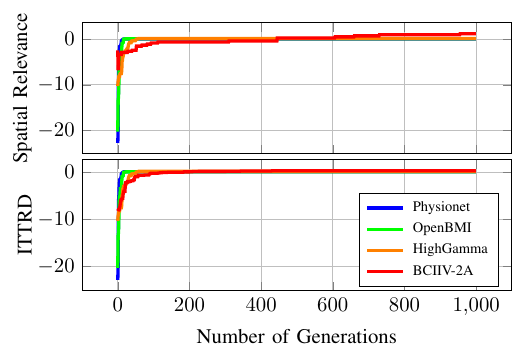}
    \end{subfigure}
    \caption{Objective-wise convergence behaviour of NSGA-II, MOPSO, and  MOEA/D algorithms on EEG datasets (Left to right).}
    \label{fig:convergence_all}
\end{figure*}

\subsection{Hyper-Parameter Settings}

The evaluation of the channel selection framework was conducted using these three multi-objective optimisation algorithms: NSGA-II, MOPSO, and MOEA/D.  For the NSGA-II algorithm, a population size of 10 was used over 1000 generations, with crossover and mutation probabilities set to 0.7 and 0.1, respectively. In contrast, the MOPSO algorithm operated with a swarm size of 10 for 100 iterations, maintaining the inertia weight ($\boldsymbol{\omega}$) at 0.5 and setting both acceleration coefficients to 2 ($c_1 = c_2 = 2$). To enhance diversity among the Pareto-optimal solutions, a repository of 100 solutions and 10 grid divisions was utilised. The MOEA/D algorithm used decomposition with 19 uniformly distributed weight vectors and a neighbourhood size of 10, with mating restricted to neighbours with a probability of 0.7. This algorithm was run for 1000 generations, with crossover and mutation probabilities set to 0.7 and 0.1, respectively.

\section{Results}

The proposed framework is evaluated on EEG datasets, and accuracy comparisons are made among NSGA-II, MOPSO, MOEA/D, and greedy based on the average number of channels utilised, as well as on the selected and total number of features. The convergence properties of each algorithm are examined, along with an analysis of the structure of Pareto-optimal solutions and the spatial clustering of selected channels. Finally, a statistical analysis is performed to confirm the significance of the differences observed between the algorithms.

\subsection{Performance of Channel Selection Approaches}

The classification accuracy evaluation for EEG datasets using all channel selection methods is shown in Table~\ref{tab:with_domain_knowledge}. For each subject, the three MOO algorithms (i.e., NSGA-II, MOPSO, MOEA/D) generated ten candidate optimal channel subsets. Among these candidate subsets, the solution containing either C3 or C4 and yielding the highest classification accuracy after feature selection was selected for evaluation. The average classification accuracy across all subjects was then computed using both the complete and selected feature sets, along with the average number of selected channels under a maximum channel constraint of 16. 

Across all datasets, using selected features consistently improved classification performance compared to using the full feature set. On the Physionet dataset, MOPSO achieved the highest improvement, increasing classification accuracy from 76\% using all features to 87\% using selected features with 16 selected channels. Similarly, NSGA-II and MOEA/D improved the accuracy from 73\% to 83\% and from 71\% to 80\%, respectively. The greedy baseline also showed a smaller improvement, from 91\% to 93\%, using 11.77 selected channels.

On the BCIIV-2A dataset, all optimisation frameworks showed noticeable performance gains after feature selection. NSGA-II improved classification accuracy from 52\% to 61\%, while MOPSO and MOEA/D increased performance from 53\% to 63\% and from 54\% to 63\%, respectively. The greedy baseline improved from 52\% to 65\% using only 4.50 selected channels. A similar trend was observed on the HighGamma dataset, where NSGA-II improved accuracy from 70\% to 75\%, and MOPSO improved from 70\% to 74\%. MOEA/D also demonstrated improvement from 64\% to 69\% after feature selection. Although the greedy baseline increased from 67\% to 71\% using fewer selected channels, the selected subsets were less consistently concentrated within physiologically relevant sensorimotor cortex regions.

On the OpenBMI dataset, MOPSO achieved the greatest improvement in classification accuracy, increasing it from 61\% to 71\% using selected features. Similarly, NSGA-II improved performance from 61\% to 67\%, while the greedy baseline improved from 62\% to 69\%. In contrast, MOEA/D showed comparatively limited improvement, achieving 54\% accuracy using selected features. We observed that reducing the number of features improves classification performance while reducing computational complexity and maintaining physiologically meaningful channel subsets.

\subsection{Convergence Characteristics of Channel Selection Algorithms in MOO Framework}

The convergence behaviour of multi-objective optimisation algorithms across different datasets, concerning both objectives, is illustrated in Fig~\ref{fig:convergence_all}. All three algorithms displayed significant improvements early in the optimisation process, followed by stabilisation as they approached Pareto-optimal channel subsets.

For the spatial relevance objective, NSGA-II converged between 200 and 300 generations, with objective values stabilising around 8-10 on Physionet and BCIIV-2A, and between 5-7 on OpenBMI and HighGamma. MOEA/D converged rapidly in the first 100-200 generations and stabilised around 16-18 for Physionet and BCIIV-2A, while  OpenBMI and HighGamma converged after approximately 300-400 generations and stabilised between 10-12. In contrast, MOPSO converged considerably earlier, with Spatial Relevance values stabilising around $-0.60$ to $-0.62$ on Physionet and OpenBMI, $-0.64$ to $-0.66$ on HighGamma, and $-0.70$ to $-0.72$ on BCIIV-2A. We found that Physionet and BCIIV-2A consistently exhibit stronger spatial localisation of MI-related activity compared with OpenBMI and HighGamma.

For the ITTRD objective, NSGA-II demonstrated relatively smooth convergence behaviour, reaching stable values between 400 and 500 generations across all datasets. Physionet converged between approximately $-50$ and $-80$, BCIIV-2A between $-80$ and $-120$, whereas OpenBMI and HighGamma converged below $-150$, indicating stronger task-related desynchronisation patterns in these datasets. MOEA/D exhibited slower but stable convergence behaviour for ITTRD, converging around 400-450 generations on Physionet with values between 17 and 18, around 500 generations on OpenBMI with values between 14 and 15, and approximately 600 generations on HighGamma and BCIIV-2A with lower values between 10 and 12. In comparison, MOPSO showed dataset-dependent oscillatory behaviour for ITTRD. 

On Physionet, HighGamma, and BCIIV-2A datasets, the objective values remained nearly constant around zero throughout optimisation. This is because the global best ($gbest$) solutions for ITTRD were already discovered in the initial iterations, after which the swarm failed to generate improved solutions, resulting in constant objective values. In these cases, the repository is dominated by early solutions, and particle updates no longer affect the ITTRD objective. For the OpenBMI dataset, on the other hand, the objective values exhibit large oscillations, with ITTRD dropping to extreme negative values ($-800$) before returning to feasible regions near zero. This instability likely arises from the interaction between binary position updates and the sigmoid transformation, compounded by the imbalance in objective scales: ITTRD has a much larger dynamic range than spatial relevance, leading the swarm to overemphasise ITTRD updates and occasionally generate degenerate solutions.

The convergence behaviour observed across the three optimisation algorithms highlights the influence of optimisation strategy on the search for optimal channel subsets. NSGA-II demonstrated stable convergence while maintaining solution diversity through non-dominated sorting and crowding-distance preservation. MOPSO showed faster convergence during the initial iterations; however, its performance was more sensitive to objective-scale imbalance, leading to oscillatory behaviour and premature fixation on the global best for the ITTRD objective on certain datasets. In comparison, MOEA/D demonstrated stronger exploitation through decomposition-based optimisation and achieved stable convergence across datasets, although more complex temporal desynchronisation patterns required longer optimisation times.

\begin{table*}
\centering
\caption{Top 16 selected channels using the multi-objective optimisation framework on EEG datasets.}
\label{tab:Top_16_selected}
\begin{tabular}{@{}c|c|l@{}}
\toprule
\textbf{Algorithm} & \textbf{Datasets} & \textbf{Selected Channels} \\ \midrule
\multicolumn{1}{c|}{\multirow{4}{*}{NSGA-II}} & \multicolumn{1}{c|}{Physionet} & Cz, FCz, CPz, C4, Fz, C6, FC6, FC4, Pz, C3, CP4, C5, T8, CP6, FC3, CP5 \\ \cmidrule(l){2-3}
\multicolumn{1}{c|}{} & \multicolumn{1}{c|}{OpenBMI} & Cz, C4, Pz, CPz, C3, Fz, CP6, CP5, CP3, CP4, FC4, FC3, C5, C6, C2, FC6 \\ \cmidrule(l){2-3}
\multicolumn{1}{c|}{} & \multicolumn{1}{c|}{HighGamma} & CCP1h, FCC2h, CCP2h, Cz, FCC1h, FCz, CPz, FFC1h, C2, Fz, Pz, C1, CCP4h, FC2, FCC3h, FCC4h \\ \cmidrule(l){2-3}
\multicolumn{1}{c|}{} & \multicolumn{1}{c|}{BCIIV-2A} & FCz, Fz, C5, Cz, FC3, Pz, C2, C3, C4, C6, CP3, CPz, C1, CP4, FC4, CP2 \\ \midrule

\multicolumn{1}{c|}{\multirow{4}{*}{MOPSO}} & \multicolumn{1}{c|}{Physionet} & C3, C4, Cz, FCz, Pz, CPz, Fz, C6, FC4, FC6, C5, CP6, CP4, C2, FC3, FC5 \\ \cmidrule(l){2-3} 
\multicolumn{1}{c|}{} & \multicolumn{1}{c|}{OpenBMI} & Fz, Cz, CPz, Pz, C3, CP3, CP5, FC4, C4, CP4, FC1, FC3, FC5, C5, C6, TTP7h \\ \cmidrule(l){2-3} 
\multicolumn{1}{c|}{} & \multicolumn{1}{c|}{HighGamma} & CCP1h, CCP2h, Cz, FCC1h, FCC2h, FCz, FFC1h, CPz, C4, Pz, FCC3h, Fz, CCP6h, FCC4h, C3, CCP3h \\ \cmidrule(l){2-3} 
\multicolumn{1}{c|}{} & \multicolumn{1}{c|}{BCIIV-2A} & Cz, C5, FC3, FCz, Fz, CPz, C3, Pz, CP3, C1, CP2, C2, C4, FC1, FC2, FC4 \\ \midrule

\multicolumn{1}{c|}{\multirow{4}{*}{MOEA/D}} & \multicolumn{1}{c|}{Physionet} & C4, T8, CPz, C3, FC6, FCz, TP8, FT7, CP6, AF7, C6, FT8, FC2, Fz, Pz, FC5 \\ \cmidrule(l){2-3} 
\multicolumn{1}{c|}{} & \multicolumn{1}{c|}{OpenBMI} & Cz, FC1, Fz, T8, P8, Pz, CP4, FC2, FC6, FC4, C5, TPP8h, C3, TP7, CP3, P4 \\ \cmidrule(l){2-3}
\multicolumn{1}{c|}{} & \multicolumn{1}{c|}{HighGamma} & CCP1h, CPz, Cz, FCC1h, FCC2h, CCP2h, FCz, C2, C1, FFC1h, FC2, CP4, FCC3h, Fz, FCC4h, Pz \\ \cmidrule(l){2-3}
\multicolumn{1}{c|}{} & \multicolumn{1}{c|}{BCIIV-2A}  & C5, CPz, Cz, C3, FCz, Fz, C6, CP3, FC3, Pz, CP1, FC1, FC4, C4, C2, CP2 \\ \bottomrule
\end{tabular}
\end{table*}

\begin{figure}[!t]
    \centering
    \includegraphics[width=\linewidth]{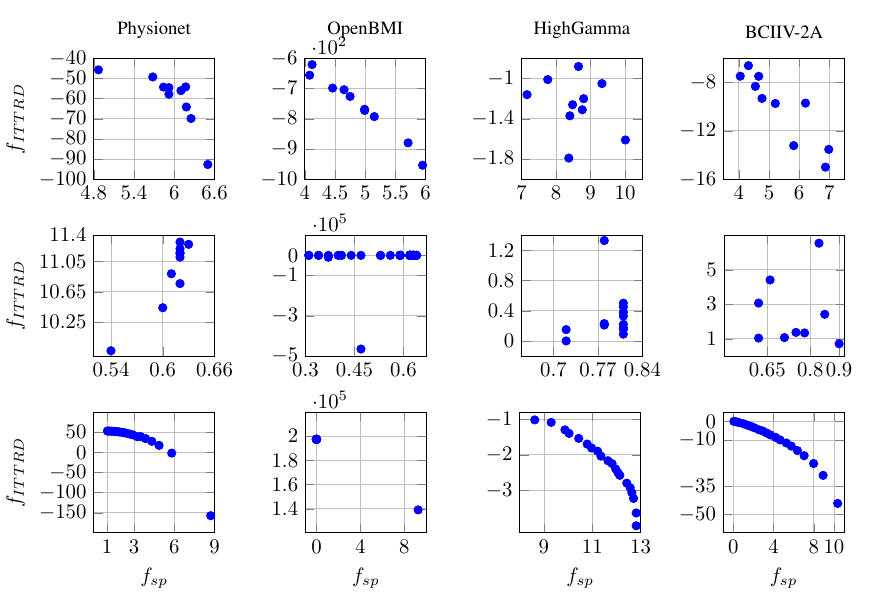}
    \caption{Average Pareto optimal solutions for datasets: Top row - NSGA-II; Middle row - MOPSO; Bottom row - MOEA/D.}
    \label{Fig:Avg_Pareto}
\end{figure}

\begin{figure*}[htbp]
    \centering
    \begin{subfigure}{0.20\textwidth}
        \includegraphics[width=\linewidth]{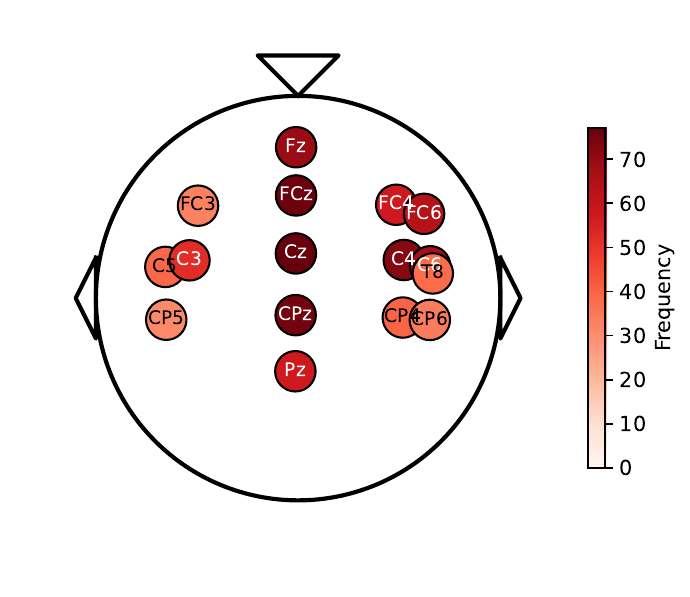}
    \end{subfigure}
    \begin{subfigure}{0.20\textwidth}
        \includegraphics[width=\linewidth]{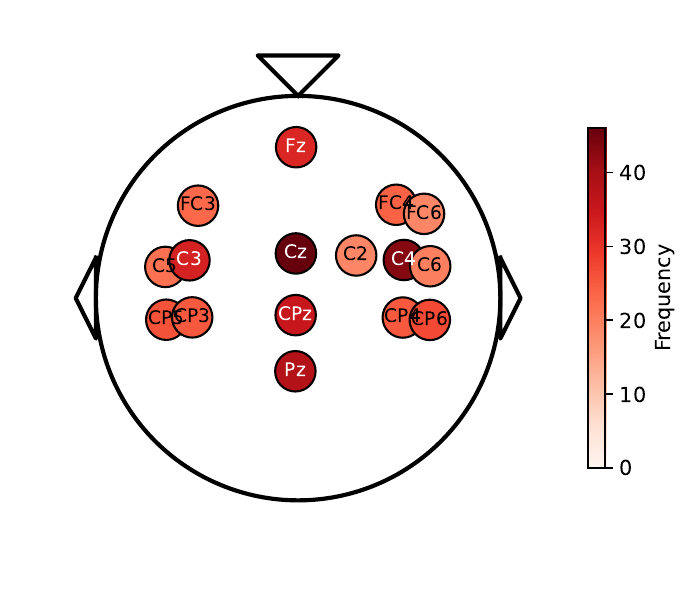}
    \end{subfigure}
    \begin{subfigure}{0.20\textwidth}
        \includegraphics[width=\linewidth]{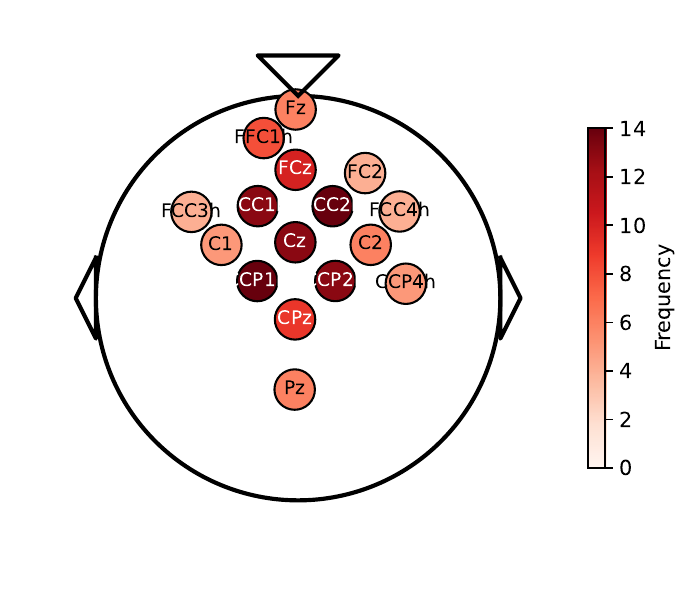}
    \end{subfigure}
    \begin{subfigure}{0.20\textwidth}
        \includegraphics[width=\linewidth]{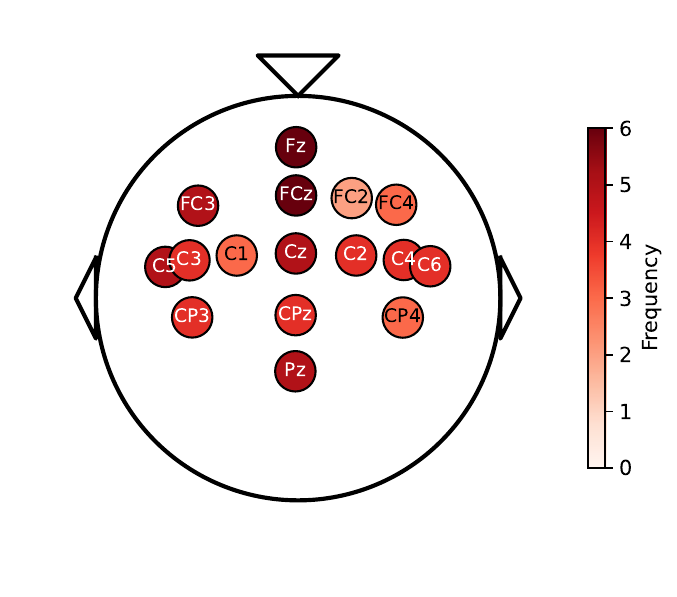}
    \end{subfigure}
    \\[-0.5cm]
    \begin{subfigure}{0.20\textwidth}
        \includegraphics[width=\linewidth]{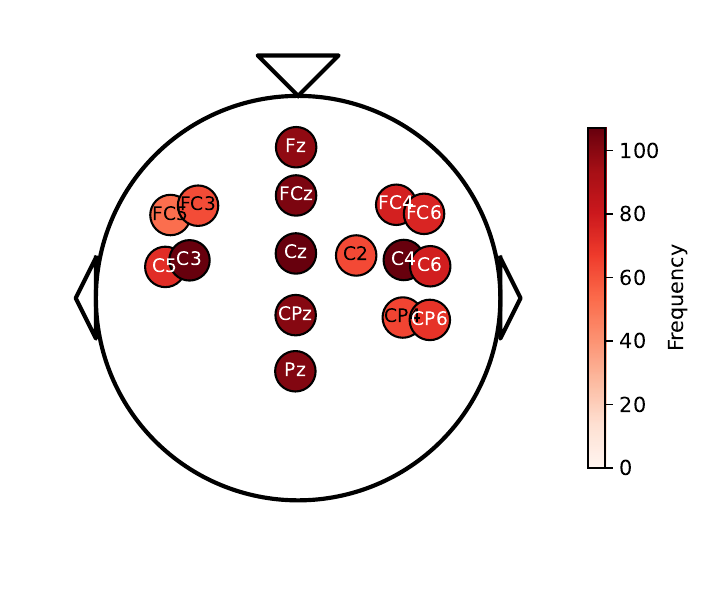}
    \end{subfigure}
    \begin{subfigure}{0.20\textwidth}
        \includegraphics[width=\linewidth]{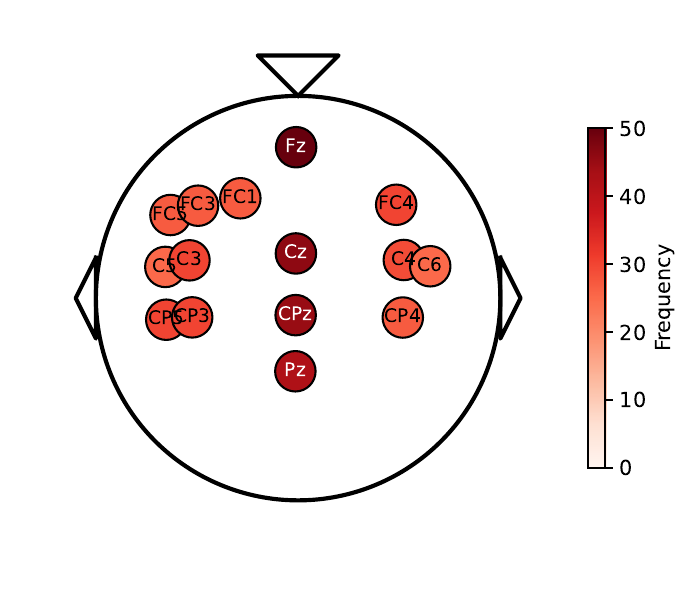}
    \end{subfigure}
    \begin{subfigure}{0.20\textwidth}
        \includegraphics[width=\linewidth]{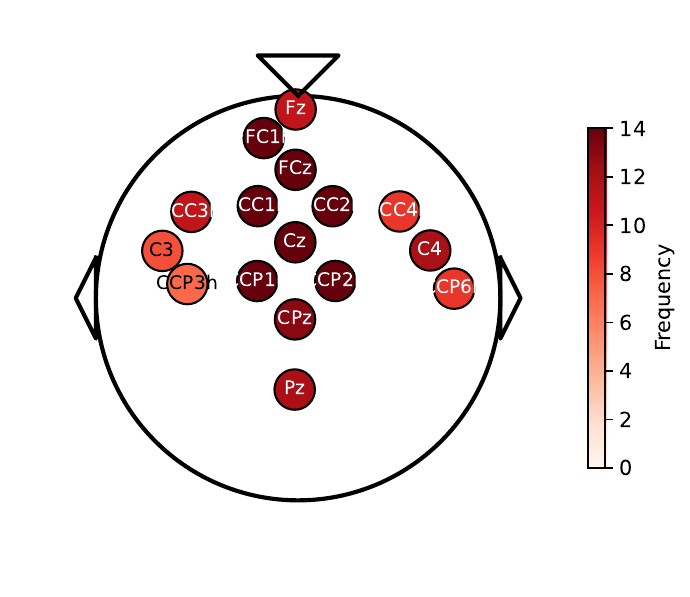}
    \end{subfigure}
    \begin{subfigure}{0.20\textwidth}
        \includegraphics[width=\linewidth]{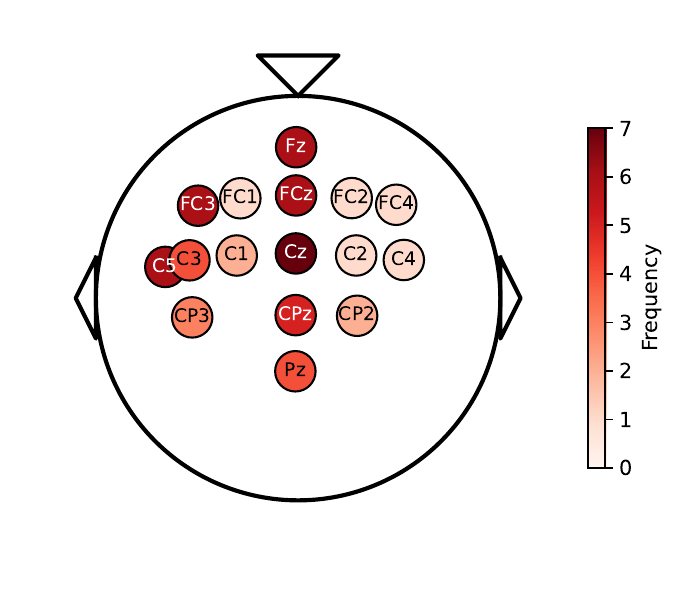}
    \end{subfigure}
    \\[-0.5cm]
    \begin{subfigure}{0.20\textwidth}
        \includegraphics[width=\linewidth]{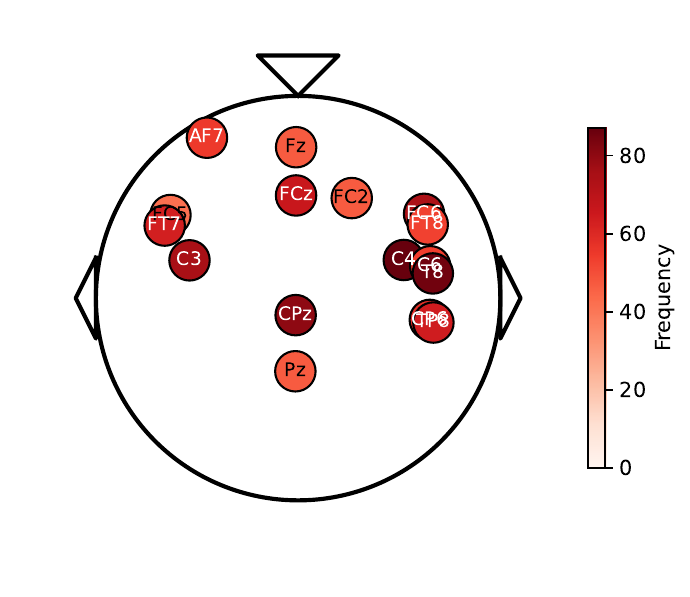}
    \end{subfigure}
    \begin{subfigure}{0.20\textwidth}
        \includegraphics[width=\linewidth]{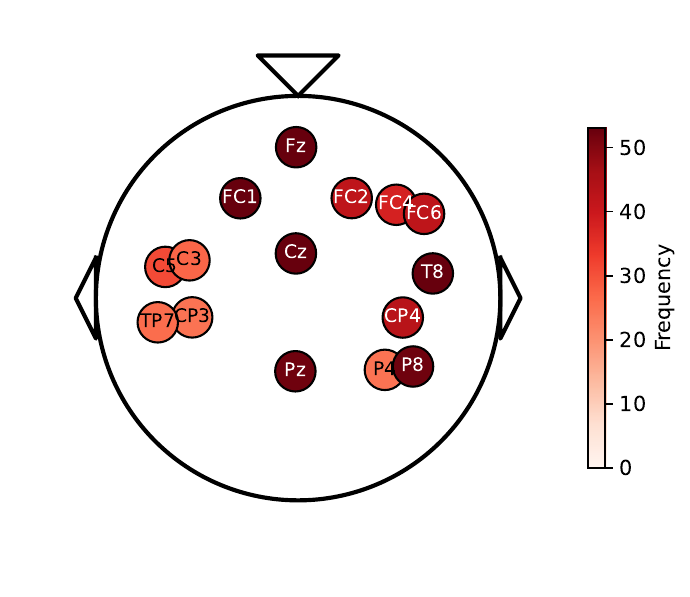}
    \end{subfigure}
    \begin{subfigure}{0.20\textwidth}
        \includegraphics[width=\linewidth]{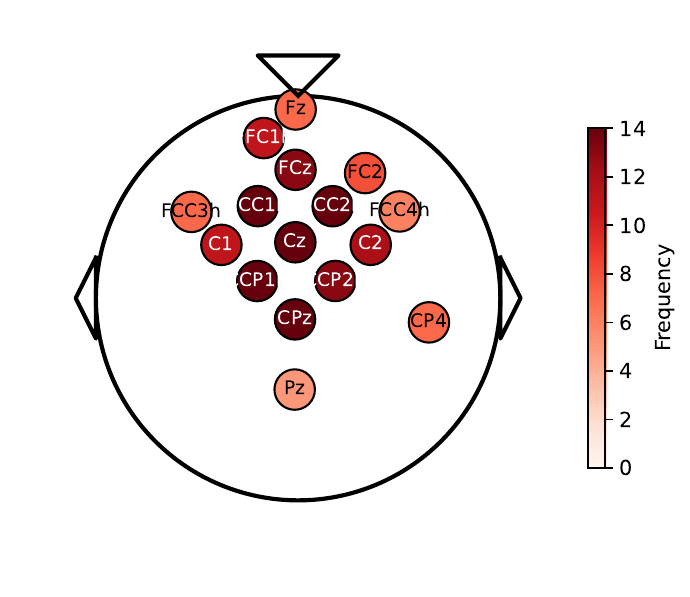}
    \end{subfigure}
    \begin{subfigure}{0.20\textwidth}
        \includegraphics[width=\linewidth]{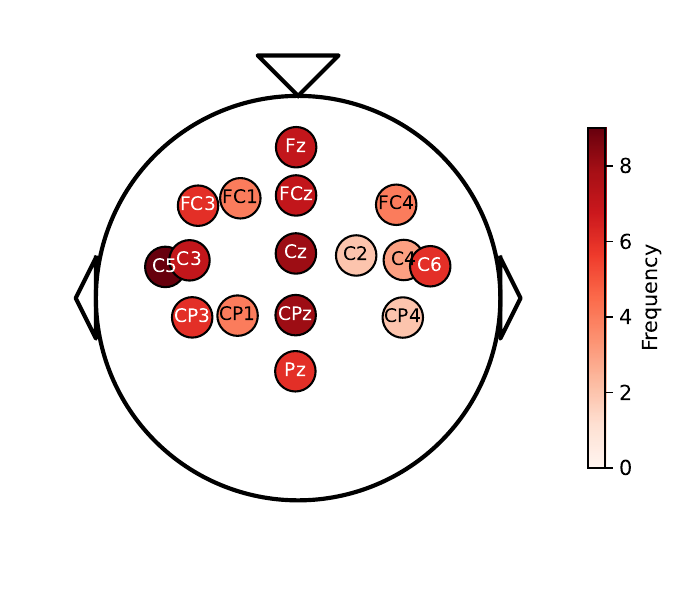}
    \end{subfigure}
    \\[-0.5cm]
    \caption{Topomaps for Physionet, OpenBMI, HighGamma, and BCIIV-2A (left to right), showing NSGA-II (top), MOPSO (middle), and MOEA/D (bottom).}
    \label{fig:all_topomaps_grid}
\end{figure*}

\subsection{Pareto-front \& Spatial Distribution of Selected Channels}

A key aspect of multi-objective optimisation is the characterisation of the Pareto-front, which represents the set of non-dominated solutions that capture the trade-off between competing objectives~\cite{multi_objective_book}. With both $f_{ITTRD}$  and $f_{sp}$ maximised simultaneously, the Pareto front reveals how well different algorithms balance these objectives; ideally, it curves towards the top-right corner of the objective space, indicating high performance on both criteria. To explore this trade-off, we computed the average Pareto-optimal solutions for each algorithm by averaging the \(k\)-th solution (\(k = 1, \ldots, 10\)) across all subjects within each of the four datasets. The averaged Pareto-optimal solutions are shown in Fig.~\ref{Fig:Avg_Pareto}, providing a comparative view of the algorithms across datasets in terms of their ability to approximate the optimal trade-offs.

The three algorithms exhibit distinct characteristics in approximating the Pareto-front across datasets (see Fig.~\ref{Fig:Avg_Pareto}). NSGA-II produces well-structured trade-offs with a good spread of non-dominated solutions, indicating effective diversity preservation through its dominance-based sorting and crowding distance mechanism. Although the averaged solutions for MOPSO appear scattered, subject-level analysis reveals that the algorithm often generates Pareto-fronts with the expected curvature. This variability reflects the stochastic nature of swarm dynamics, which enables the identification of extreme trade-offs but can reduce uniformity when aggregated. By contrast, MOEA/D consistently yields smooth and coherent fronts, particularly evident in the HighGamma and BCIIV-2A datasets, where solutions follow continuous curvatures toward 
the ideal region. 

While Pareto-fronts provide an overview of the trade-offs achieved by different algorithms, they do not by themselves reveal the specific spatial patterns underlying the selected solutions. To gain further insight, we further investigate the channel subsets corresponding to these fronts. The final optimal channel subset comprises the channels listed in Table~\ref{tab:Top_16_selected}. For better visualisation of the clusters formed by the obtained channel subsets, topomaps are plotted for each algorithm across the four datasets as shown in Fig.~\ref{fig:all_topomaps_grid}. Each map highlights selected channels with varying frequency intensities, represented by a colour gradient: red indicates higher frequency activity, and lighter shades denote lower activity. The colour bar shown in Fig.~\ref{fig:all_topomaps_grid} is mapped to the number of subjects in a dataset. A consistent observation across all methods is the frequent selection of channels located over the motor cortex regions, particularly around C3 and C4, which are well-established as discriminative areas for MI tasks. NSGA-II shows a strong concentration near central electrodes, indicating a focused selection of MI task-specific regions. In contrast, MOPSO demonstrates a broader distribution, often including frontal and parietal electrodes, suggesting that it captures visual contextual information beyond the motor cortex. MOEA/D provides a middle ground, with selections spanning both central and neighbouring areas, thereby balancing focus and diversity. 

\subsection{Statistical Evaluation of Channel Selection Approaches}

A one-way ANOVA was performed to compare the performance of NSGA-II, MOPSO, MOEA/D, and the Greedy channel selection method across four datasets. Statistically significant differences were observed for the Physionet dataset (F = 20.25, $p < 0.001$), BCIIV-2A dataset (F = 9.20, $p < 0.001$), and OpenBMI dataset (F = 21.96, $p < 0.001$). In contrast, the HighGamma dataset showed no significant differences among the methods (F = 1.56, $p = 0.21$), indicating comparable performance across all algorithms. We observed that, across most datasets, classification performance varies substantially with the chosen channel selection strategy.

To identify the methods contributing to these differences, a Tukey HSD post-hoc analysis was performed. NSGA-II significantly outperformed the Greedy method on the Physionet, BCIIV-2A, and OpenBMI datasets ($p < 0.01$). Similarly, MOPSO and MOEA/D achieved significantly higher accuracies than the Greedy baseline on these datasets ($p < 0.05$). No statistically significant differences were observed among the three MOO algorithms, indicating comparable performance among the multi-objective optimisation approaches. Consistent with the earlier ANOVA results, none of the pairwise comparisons was significant for the HighGamma dataset.

A combined one-way ANOVA was further performed by pooling the results from all datasets and subjects. The analysis revealed a highly significant difference in classification performance among the algorithms (F = 66.07, $p < 4.9 \times 10^{-39}$). This shows that channel selection strategy plays an important role in classification performance, while also highlighting the effectiveness of multi-objective optimisation approaches over the greedy baseline.


\section{Discussion}

This work presents a domain-informed multi-objective EEG channel selection framework that utilises evolutionary optimisation algorithms, including NSGA-II, MOPSO, and MOEA/D. Specifically, the framework integrates two objective functions to optimise spatial weighting and functional discriminability simultaneously by incorporating motor cortex priors into the optimisation process. It successfully identified channel subsets concentrated in the sensorimotor cortex across four datasets. In addition to uncovering spatial patterns, the Pareto-front structure of the identified solutions demonstrated effective trade-offs between the spatial relevance of both objectives and ITTRD (intratrial task-related desynchronisation). This highlights the framework's ability to obtain compact, relevant channel subsets for motor imagery (MI) activity. Although the greedy approach achieved higher accuracy in some datasets, the resulting channel subsets lacked neurophysiological interpretability. The selected electrodes were scattered across the scalp without a consistent spatial pattern and often included channels outside the sensorimotor cortex. This reflects the tendency of purely accuracy-driven methods to exploit cues specific to classifiers rather than focusing on physiologically meaningful signals.

The proposed framework achieved consistent performance gains across multiple MI datasets. On the Physionet dataset, the proposed MOPSO-based framework achieved 87\% classification accuracy using 16 selected channels. This represents a substantial improvement over the IMOCS framework proposed by Handiru et al.~\cite{Vikram}, which reported 79.5\% accuracy using the same no of channels. The improvement suggests that preserving objective independence through Pareto-based optimisation enables more effective identification of channel subsets that are both spatially relevant and functionally discriminative. In addition, the proposed framework was evaluated on a larger cohort of 109 subjects, demonstrating improved robustness and scalability.

Moreover, a domain-informed multi-objective EEG channel selection framework demonstrated consistent performance across OpenBMI, HighGamma, and BCIIV-2A, highlighting its robustness across datasets with varying channel densities and recording protocols. On the OpenBMI, the proposed framework achieved classification accuracies between 67-71\% using only 10-16 selected channels, providing an effective balance between decoding performance and channel reduction. Similarly, on the HighGamma dataset, NSGA-II and MOPSO achieved accuracies of 74\%-75\% with compact channel subsets, indicating stable optimisation performance even in high-density EEG recordings. For the BCIIV-2A dataset, the proposed framework maintained competitive classification performance while substantially reducing the number of selected channels.

This work presents that integrating neurophysiological knowledge into the optimisation process significantly enhances the efficiency and relevance of the selected channel subsets~\cite{10781846,10650619, Alotaiby2015,11343575, KIM2023103009}. Unlike traditional methods that focus on classifier-driven objectives~\cite{Che_Yau_Kee, Moein}, our approach strategically targets channels linked to motor cortex activity, ensuring that selected channels are both statistically relevant and physiologically meaningful. This balance improves classification performance and enhances the interpretability of the results, leading to more insightful analyses in MI tasks. 


\section{Conclusion}

This work proposes a domain-informed multi-objective EEG channel selection framework for MI-based BCIs, effectively optimising Gaussian-kernel-based spatial weighting and functional discriminability through ITTRD. By employing NSGA-II, MOPSO, and MOEA/D algorithms, we identified Pareto-optimal channel subsets across various datasets, including Physionet, OpenBMI, HighGamma, and BCIIV-2A, achieving competitive classification performance while significantly reducing the number of selected channels. This work set a benchmark for multi-objective channel selection on four publicly available EEG datasets, demonstrating a robust balance of optimisation stability, channel reduction, and neurophysiological interpretability. This framework can be adapted for other EEG-based BCI paradigms, such as multi-class classification, steady-state visually evoked potential (SSVEP), and P300 systems. Future work will also explore adaptive real-time optimisation, subject-independent channel selection, and the implementation of wearable BCIs using compact, low-complexity EEG configurations.


\bibliographystyle{IEEEtran}

\bibliography{main}

\end{document}